\documentclass[12pt]{iopart}
\usepackage{graphicx}
\usepackage[usenames]{color}
\expandafter\let\csname equation*\endcsname\relax
\expandafter\let\csname endequation*\endcsname\relax
\usepackage{amssymb,amsmath}
\usepackage[square,comma,compress,numbers]{natbib}

\usepackage{dcolumn}% Align table columns on decimal point
\usepackage{bm}% bold math
\usepackage{float}

\newcommand{\eq}[1]{eq.~(\ref{#1})}
\newcommand{\fig}[1]{Fig.~\ref{#1}}

\newcommand{\gj}[6]{ \begin{pmatrix}
 #1 & #2 & #3 \\
 #4 & #5 & #6 
 \end{pmatrix}}

\begin{document}
\title{Carbon monoxide interacting with free-electron-laser pulses}
\author{H. I. B. Banks}
\address{Department of Physics and Astronomy, University College London, Gower Street, London WC1E 6BT, United Kingdom}
\author{A. Hadjipittas}
\address{Department of Physics and Astronomy, University College London, Gower Street, London WC1E 6BT, United Kingdom}
\author{A. Emmanouilidou}
\address{Department of Physics and Astronomy, University College London, Gower Street, London WC1E 6BT, United Kingdom}

\begin{abstract}
We study the interaction of a heteronuclear diatomic molecule, carbon monoxide, with a free-electron laser (FEL) pulse. We compute the ion yields and the intermediate states by which the ion yields are populated. We do so using rate equations, computing all relevant molecular and atomic photoionisation cross-sections and Auger rates. We find that the charge distribution of the carbon and oxygen ion yields differ. By varying the photon energy, we demonstrate how to control higher-charged states being populated mostly by carbon or oxygen. Moreover, we identify the differences in the resulting ion yields and pathways populating these yields between a homonuclear molecule, molecular nitrogen, and a heteronuclear molecule, carbon monoxide, interacting with an FEL pulse. These two molecules have similar electronic structure. We also identify the proportion of each ion yield which accesses a two-site double-core-hole state and tailor pulse parameters to maximise this proportion.
 \end{abstract}
\date{\today}
\submitto{\jpb}
\maketitle
 
 %\ioptwocol
%%=========================================================================
\section{Introduction}

X-ray free-electron lasers (XFELs) \cite{FELhistory,Bostedt2013,Emma2010} have introduced new tools and techniques for the investigation and imaging of atoms and molecules \cite{Marangos2011,Ullrich2012}. The x-ray energy of the photons and high intensity of the FEL pulses allow for high-resolution images of large biomolecules \cite{Schlichting2012,Neutze2000, Redecke227}. These x-ray photons are also more likely to ionise inner-shell electrons than valence electrons, resulting in the creation of core holes. If another core hole is created, before the atom or molecule has time to relax via Auger processes, a double-core-hole (DCH) state is formed. In molecules, there are two types of DCH states, those where both core holes are on the same atomic site, i.e. single-site double-core-hole (SSDCH) states, and those where the core holes are on different atomic sites, i.e. two-site double-core-hole (TSDCH) states. These TSDCH states are particularly interesting due to their sensitivity to their chemical environment \cite{Cederbaum1986, Cederbaum1987, Tashiro2010, Salen2012, Piancastelli2013}. Double-core-hole states are short-lived, as they decay via Auger processes. This process involves a core hole being filled in by a valence electron, while the released energy results in the ejection of another valence electron. There has been a significant amount of both experimental \cite{Berrah2011, Santra2009, Fang2010, Fang2012, Zhaunerchyk2015} and theoretical \cite{Buth2012, Liu2016} work regarding these states.

We investigate the influence of two-site double-core-hole states during the interaction of an FEL pulse with a heteronuclear diatomic molecule. Specifically, we identify the set of pulse parameters that maximise the production of two-site double-core-hole states during the interaction of carbon monoxide (CO) with an FEL pulse. The interaction of CO with an FEL pulse has been the focus of several studies, as it pertains to biomolecules \cite{Hill1994,Levantino2015}. The formation of TSDCH states in $\rm CO$ has been detected experimentally through photoelectron spectra \cite{Berrah2011}, with supporting theoretical energy calculations \cite{Tashiro2010}. In the current study, we investigate the effect of the photon energy, pulse duration and intensity of the pulse on the contribution of TSDCHs in the formation of the final ion yields. As a result, we identify the most appropriate FEL pulse parameters for maximising the proportion of the final ion yields that accesses a TSDCH. In addition, we investigate the role of TSDCH states during the interaction of an FEL pulse with a homonuclear versus a heteronuclear diatomic molecule. We do so in the context of molecular nitrogen ($\rm N_2$) versus $\rm CO$, two diatomic molecules of similar electronic structure. 

Moreover, we investigate whether carbon or oxygen populate more highly-charged states during the interaction of $\rm CO$ with an FEL pulse. That is, by varying the photon energy as well as the duration and intensity of the FEL pulse we identify which pulse parameters favour higher-charged states being populated by carbon or oxygen. In addition, for the same FEL pulse parameters we compare the resulting carbon and oxygen ion yields with the atomic nitrogen yields resulting from the FEL pulses interacting with $\rm N_2$. This allows us to identify additional differences in the interaction of homonuclear and heteronuclear molecules with FEL pulses.

%%=========================================================================

\section{Method}
\label{Method}

\subsection{Rate Equations}

We use rate equations to model the interaction of FEL pulses with $\rm CO$. Every energetically accessible state of $\rm CO$ is denoted by its electronic configuration, ($1\sigma^{a},2\sigma^{b},3\sigma^{c},4\sigma^{d},1\pi_{x}^{e},1\pi_{y}^{f}, 5\sigma^{g}$), where $a,b,c,d,e,f$ and $g$ are the number of electrons occupying a molecular orbital. This occupancy is 0, 1 or 2. These rate equations were discussed in detail in previous work in the context of the homonuclear molecule $\rm N_2$ ($1\sigma_g^{a},1\sigma_u^{b},2\sigma_g^{c},2\sigma_u^{d},1\pi_{ux}^{e},1\pi_{uy}^{f}, 3\sigma_g^{g}$) interacting with FEL pulses \cite{Banks2017}. 
In the rate equations describing the interaction of $\rm CO$ with an FEL pulse we employ the single-photon ionisation cross-section and Auger rates for all energetically allowed transitions in $\rm CO$ as well as its atomic fragments. Atomic units are used in this work, unless otherwise stated. To obtain these cross sections and rates we need to compute the bound and the continuum atomic and molecular orbitals for all energetically accessible states. To simplify the computations involved for the photoionisation cross-section and Auger rates, we express these orbitals in terms of a single-centre expansion (SCE) \cite{Demekhin2011}. 

\subsection{Bound and continuum orbitals}

We obtain the molecular and atomic bound wavefunctions $\psi_i$ using the Hartree-Fock technique in Molpro \cite{Molpro}, a quantum chemistry package. We use the cc-pVTZ basis set to obtain the bound wavefunctions for $\rm CO$ with the nuclei fixed at an equilibrium distance of 1.128 \AA \cite{Gilliam1950}. In the single-centre expansion the bound and continuum wavefunctions, $\psi_i$ and $\psi_\epsilon$ respectively, are expressed as \cite{Demekhin2011}
\begin{equation}
\psi_{(i/\epsilon)}(\textbf{r})=\sum_{lm}\frac{P^{i/\epsilon}_{lm}(r)Y_{lm}(\theta,\phi)}{r},
\label{SCE}
\end{equation}
\noindent with $\textbf{r}=(r,\theta,\phi)$ denoting the position of the electron, with respect to the centre of mass of the molecule. We denote by $Y_{lm}$, a spherical harmonic with quantum numbers $l$ and $m$ while $P^{i/\epsilon}_{lm}(r)$ denotes the single centre expansion coefficients for the orbital $i/\epsilon$. The index $i$ refers to the $i$th bound orbital, while $\epsilon$ refers to a continuum orbital with energy $\epsilon$. Since CO is a linear molecule, it has rotational symmetry and hence only one $m$ value is involved in the summation in \eq{SCE}. For a heteronuclear molecule, like CO, the wavefunctions have no gerade or ungerade symmetry and therefore both even and odd values of $l$ have to be included in \eq{SCE}. This is unlike a homonuclear molecule, like $\rm N_2$, where only odd or even values of $l$ are included in the summation in \eq{SCE} depending on whether the wavefunction has gerade or ungerade symmetry.

The continuum wavefunctions, $\psi_\epsilon$, are calculated by solving the following Hartree-Fock equations \cite{Banks2017,Demekhin2011, Bransden2003}:
%\begin{widetext}
\begin{align}
\label{eqn:schr1}
\underbrace{-\frac{1}{2} {\nabla}^2 \psi_{\epsilon}({\bf r})}_{\text{Kinetic energy}} + \underbrace{\sum_{n}^{nuc.} \frac{-Z_n}{|{\bf r}-{{\bf R}_n}|}\psi_{\epsilon}({\bf r})}_{\text{Electron-nuclei}}
+\underbrace{\sum_i^{orb.} a_i \int d{\bf r}p \frac{\psi^*_{i}({\bf r}p)\psi_{i}({\bf r}p)}{|{\bf r} - {\bf r}p|} \psi_{\epsilon}({\bf r})}_{\text{Direct interaction}}&\nonumber
\\
 - \underbrace{\sum_{i}^{orb.} b_i \int d{\bf r}p \frac{\psi^{*}_{i}({\bf r}p)\psi_{\epsilon}({\bf r}p)}{|{\bf r} - {\bf r}p|}\psi_{i}({\bf r})}_{\text{Exchange interaction}}& = \epsilon\psi_{\epsilon}({\bf r}),
\end{align} 
%\end{widetext}
\noindent where $\psi_i$ is the wavefunction of the $i$th bound molecular orbital, $a_i$ is the occupation of orbital $i$ and $b_i$ is a coefficient associated with the $i$th orbital, whose values are determined by the symmetry of the state. For more details, see our previous work \cite{Banks2017}. ${\bf R}_n$ and $Z_n$ are the position with respect to the centre of mass, and the charge of nucleus $n$, respectively. In order to compute the continuum orbitals $\psi_\epsilon$, we substitute $\psi_\epsilon$ and $\psi_i$ using \eq{SCE} and use the non-iterative method \cite{Banks2017,Demekhin2011} to solve for the $P^\epsilon_{lm}$ coefficients.

\subsection{Photo-ionisation}

The photoionisation cross-section for an electron transitioning from an initial molecular orbital $\psi_{i}$ to a final continuum molecular orbital $\psi_{\epsilon}$ is given by \cite{ModernQM} 
 \begin{equation}
\sigma_{i\rightarrow \epsilon} = \frac{4}{3}\alpha\pi^2 \omega N_i \sum_{M=-1,0,1} {\left|D_{i\epsilon}^M\right|}^{2}.
\end{equation}
\noindent The fine-structure constant is denoted by $\alpha$, the photon energy by $\omega$, the occupation number of orbital $i$ by $N_i$ and the magnetic quantum number of the photon by $M$. In the length gauge, using \eq{SCE}, the matrix element $D_{i\epsilon}^M$ is given by
\begin{flalign}
\label{eq:pi_wig}
D_{i\epsilon}^M& = \sqrt{\frac{4\pi}{3}}\sum_{lm,l'm'}\int^\infty_0 dr P^{i*}_{lm}(r) r P^\epsilon_{l'm'}(r)\times \int d\Omega Y^{*}_{lm}(\theta,\phi)Y_{l'm'}(\theta,\phi)Y_{1M}(\theta,\phi)&&\\\nonumber
 & = \sum_{lm,l'm'}{(-1)}^{m} \sqrt{(2l+1)(2l'+1)} 
 \begin{pmatrix} 
l & l' & 1 \\ 
0 & 0 & 0
\end{pmatrix} \begin{pmatrix}
l & l' & 1 \\ 
-m & m' & M
\end{pmatrix} \int^\infty_0 dr P^{i*}_{lm}(r) r P^\epsilon_{l'm'}(r).&&
\end{flalign}
\noindent Eq.(4)\ clearly shows that, by adapting the SCE for the bound and continuum wavefunctions, we significantly simplify the computation of the cross-section. Namely, the result of the angular integrals is expressed in terms of the Wigner-3j symbols \cite{AtoSpec} and we only have to solve a 1D integral numerically, which involves the single-centre expansion coefficients, $P^{i/\epsilon}_{lm}$. The computation of the matrix element $D_{i\epsilon}^M$ is more intensive for the heteronuclear molecule, CO, than the homonuclear $\rm N_{2}$ as it involves both odd and even values for the $l$ and $l'$ numbers.

\subsection{Auger decay}

Auger decay rates have been calculated with a variety of different methods in existing work \cite{Wallis2014,Bhalla,Pulkkinen1996,Lablanquie2007,Son2012,Bolognesi2006,Feyer2005,Puttner2008,Iwayama2013,Storchi2008,Griffiths1991}. The general expression for the Auger rate, $\Gamma$, is given by \cite{FermiGR}:
 \begin{flalign}
\Gamma=\overline{\sum}2\pi |\mathcal{M}|^2\equiv\overline{\sum}2\pi |\langle\Psi_{fin}|H_I|\Psi_{init}\rangle|^2,
\label{GeneralAuger}
\end{flalign}
\noindent where $\mathrm{\overline{\sum}}$ denotes a summation over the final states and an average over the initial states. $\mathrm{|\Psi_{init}\rangle}$ and $\mathrm{|\Psi_{fin}\rangle}$ are the wavefunctions of all electrons in the initial and final molecular state, respectively. $H_{I}$ is the interaction Hamiltonian. In the $m_a,m_b,S,M_S$ scheme, the Auger rate is given by \cite{Banks2017}
\begin{flalign}
\Gamma_{b,a\rightarrow s,\zeta}=
\sum_{\substack{m_am_bm_sm_\zeta\\SM_SS'M'_S}}\pi N_{ab}N_h\sum_L|\mathcal{M}|^2,
\label{TotalSM_Sgeneral}
\end{flalign}
\noindent with $N_h$ being the number of holes in the orbital to be filled. $a$, $b$ refer to the valence orbitals, $s$ to the core orbital which is filled in and $\zeta$ refers to the continuum orbitals. $S$ and $M_S$ are the total spin and its orientation before the transition and $S'$ and $M'_S$ are the total spin and its orientation afterwards. As the CO orbitals have well-defined $m$, the summations over $m$ will take only a single value. $N_{ab}$ is the weighting factor related to the occupation of the valence orbitals which fill the hole given by
 \begin{align}
N_{ab}=\Bigg\{
\begin{array}{lr}
\dfrac{N_{a}N_{b}}{2\times2}\qquad \qquad & \mathrm{for\,\,different\,\, orbitals}\\[2ex]
\dfrac{N_{a}(N_{a}-1)}{2\times2\times1} &\mathrm{for\,\,same\,\, orbital.}
\end{array}&&
\end{align}
\noindent Here, $N_a$ and $N_b$ are the occupations of orbitals $a$ and $b$, respectively. The matrix element, $\mathcal{M}$, is given by
\begin{align}
\label{AugerSM_S}
&\mathcal{M}=\delta_{S}^{S'}\delta_{M_S}^{M'_S}\sqrt{(2l_s+1)(2l_a+1)(2l_\zeta +1)(2l_b+1)}&&\\\nonumber
&\times\left[\sum_{\substack{kl_\zeta l_s\\l_bl_a}}\sum_{q=-k}^{k}(-1)^{m_s+q+m_\zeta}\right.
\int dr_1\int dr_2 P^{\zeta*}_{l_\zeta m_\zeta}(r_1)P^{s*}_{l_sm_s}(r_2)\frac{r^k_<}{r^{k+1}_>}P^{b}_{l_bm_b}(r_1)P^{a}_{l_am_a}(r_2)&&\\\nonumber
&\times\gj{l_s}{k}{l_a}{0}{0}{0}\gj{l_s}{k}{l_a}{-m_s}{q}{m_a}\gj{k}{l_\zeta }{l_b}{0}{0}{0}\gj{k}{l_\zeta}{l_b}{-q}{-m_\zeta}{m_b}&&\\\nonumber
&+(-1)^{S}\sum_{\substack{kl_\zeta l_s\\l_bl_a}}\sum_{q=-k}^{k}(-1)^{m_s+q+m_\zeta}
\int dr_1\int dr_2 P^{\zeta*}_{l_\zeta m_\zeta}(r_1)P^{s*}_{l_sm_s}(r_2)\frac{r^k_<}{r^{k+1}_>}P^{a}_{l_am_a}(r_1)P^{b}_{l_bm_b}(r_2)&&\\\nonumber
&\left.\times\gj{l_s}{k}{l_b}{0}{0}{0}\gj{l_s}{k}{l_b}{-m_s}{q}{m_b}\gj{k}{l_\zeta}{l_a}{0}{0}{0}\gj{k}{l_\zeta}{l_a}{-q}{-m_\zeta}{m_a}\vphantom{\sum_{\substack{kl_\zeta l_s\\l_bl_a}}}\right]&&
\end{align}
\noindent where $r_<=min(r_1,r_2)$ and $r_>=max(r_1,r_2)$. Due to rotational symmetry, each orbital has a single well-defined $m$ number. This calculation is more computationally intensive in the heteronuclear case than in the homonuclear case, as the wavefunctions do not have gerade or ungerade symmetry and therefore will involve both odd and even values of $l$.

\subsection{Dissociation}

We treat the dissociation of the molecule phenomenologically, with rates based on FEL experiments with $\rm N_2$ \cite{Liu2016,Beylerian2004}. This approximation is justified by the similarity between $\rm CO$ and $\rm N_2$, particularly with respect to their dissociative transitions \cite{Gaire2009}. The molecular ion $\rm CO^{2+}$ is treated as dissociating with a lifetime of 100 fs \cite{Beylerian2004}. The final products of this dissociation, according to the experimental work in ref. \cite{Hitchcock1988}, are $\rm C^+$ and $\rm O^+$ with 85\% probability, $\rm C^{2+}$ and $\rm O$ with 9\% probability, and $\rm C$ and $\rm O^{2+}$ with 6\% probability. States of $\rm CO^{3+}$ without core holes are treated as instantaneously dissociating to $\rm C^+$ and $\rm O^{2+}$ or $\rm C^{2+}$ and $\rm O^{+}$ with equal probability, as is the case for $\rm N_2$ \cite{Liu2016,Banks2017}. All states of $\rm CO^{4+}$ are treated as dissociating instantaneously to $\rm C^{2+}$ and $\rm O^{2+}$ \cite{Liu2016,Banks2017}. To determine how missing molecular orbitals just before dissociation map to missing atomic orbitals just after dissociation, we calculate overlaps between molecular and atomic orbitals. Specifically, we calculate the overlap $\langle\psi_a^{\rm CO}|\psi_\alpha^{\rm C/O}\rangle$ of each molecular orbital of neutral $\rm CO$ with each atomic orbital of neutral $\rm C$ and $\rm O$, where $a$ corresponds to a molecular orbital and $\alpha$ to an atomic orbital. We find that $\langle\psi_{1\sigma}^{\rm CO}|\psi_{1s}^{\rm C}\rangle\approx1$, which means that the $1\sigma$ CO orbital with energy 542 eV corresponds to a 1s O orbital with energy 544 eV. Moreover, we find $\langle\psi_{2\sigma}^{\rm CO}|\psi_{1s}^{\rm O}\rangle\approx1$, which means that the $2\sigma$ CO orbital with energy 296 eV corresponds to a 1s C orbital with energy 297eV. The other molecular orbitals have overlaps with atomic orbitals on both atomic sites. In what follows, we use these overlaps to determine the possible dissociation products and the dissociation rates to different sets of atomic states. For example, for a $\rm CO^{2+}$ state with missing electrons in molecular orbitals $a$ and $b$, the dissociation rate to atomic states missing electrons in atomic orbitals $\alpha$ and $\beta$ is given by:
\begin{flalign}
\Gamma_{a,b\rightarrow \alpha,\beta}^{\rm C^{2+}+O}=
0.09\times\Gamma_{a,b}^{\rm CO^{2+}}\frac{\big|\langle \psi_a^{\rm CO}|\psi_\alpha^{\rm C}\rangle\big|^2\big|\langle \psi_b^{\rm CO}|\psi_\beta^{\rm C}\rangle\big|^2}{\sum_{i,j}\left|\langle \psi_a^{\rm CO}|\psi_i^{\rm C}\rangle\right|^2\left|\langle \psi_b^{\rm CO}|\psi_j^{\rm C}\rangle\right|^2}
\label{C2+Dissociation}
\end{flalign}
\begin{flalign}
\Gamma_{a,b\rightarrow \alpha,\beta}^{\rm C^{+}+O^{+}}=
0.85\times\Gamma_{a,b}^{\rm CO^{2+}}\frac{\big|\langle \psi_a^{\rm CO}|\psi_\alpha^{\rm C}\rangle\big|^2\big|\langle \psi_b^{\rm CO}|\psi_\beta^{\rm O}\rangle\big|^2}{\sum_{i,j}\left|\langle \psi_a^{\rm CO}|\psi_i^{\rm C}\rangle\right|^2\left|\langle \psi_b^{\rm CO}|\psi_j^{\rm O}\rangle\right|^2}
\label{C+O+Dissociation}
\end{flalign}
\begin{flalign}
\Gamma_{a,b\rightarrow \alpha,\beta}^{\rm C+O^{2+}}=
0.06\times\Gamma_{a,b}^{\rm CO^{2+}}\frac{\big|\langle \psi_a^{\rm CO}|\psi_\alpha^{\rm O}\rangle\big|^2\big|\langle \psi_b^{\rm CO}|\psi_\beta^{\rm O}\rangle\big|^2}{\sum_{i,j}\left|\langle \psi_a^{\rm CO}|\psi_i^{\rm O}\rangle\right|^2\left|\langle \psi_b^{\rm CO}|\psi_j^{\rm O}\rangle\right|^2},
\label{O2+Dissociation}
\end{flalign}
\noindent where $\Gamma_{a,b}^{\rm CO^{2+}}$ is the rate of dissociation of molecular ion $\rm CO^{2+}$ with electrons missing from molecular orbitals $a$ and $b$. This rate corresponds to a lifetime of 100 fs \cite{Beylerian2004}. Note, in eqs. \ref{C2+Dissociation}-\ref{O2+Dissociation}, we only consider transitions to atomic ions with electrons missing from orbitals $\alpha$ and $\beta$ if the relevant overlaps have values greater than 0.02. 

% For example, the electronic configuration $1\sigma^{1},2\sigma^{1},3\sigma^{2},4\sigma^{2},1\pi_{x}^{2},1\pi_{y}^{2}, 5\sigma^{0}$ is a state of $CO^{4+}$ and therefore will break down into $C^{2+}$ and $O^{2+}$. To determine which atomic orbitals are depopulated, we look at the overlaps of the missing molecular electrons. We note that the missing $1\sigma$ corresponds to a 1s hole on the $O$ and the missing $2\sigma$ to a 1s hole on the $C$. From the overlaps, we find that the $5\sigma$ orbital is formed from the 2p $C$ and the 2p $O$ and the charge constraints mean that one electron is missing from each. This means that $1\sigma^{1},2\sigma^{1},3\sigma^{2},4\sigma^{2},1\pi_{x}^{2},1\pi_{y}^{2}, 5\sigma^{0}$ breaks down into $1s^12s^22p^1$ carbon and $1s^12s^22p^3$ oxygen.

\section{Results}

\subsection{Ion Yields}

\subsubsection{Ion yield dependence on FEL pulse parameters}

\begin{figure}
\resizebox{\textwidth}{!}{
 \includegraphics{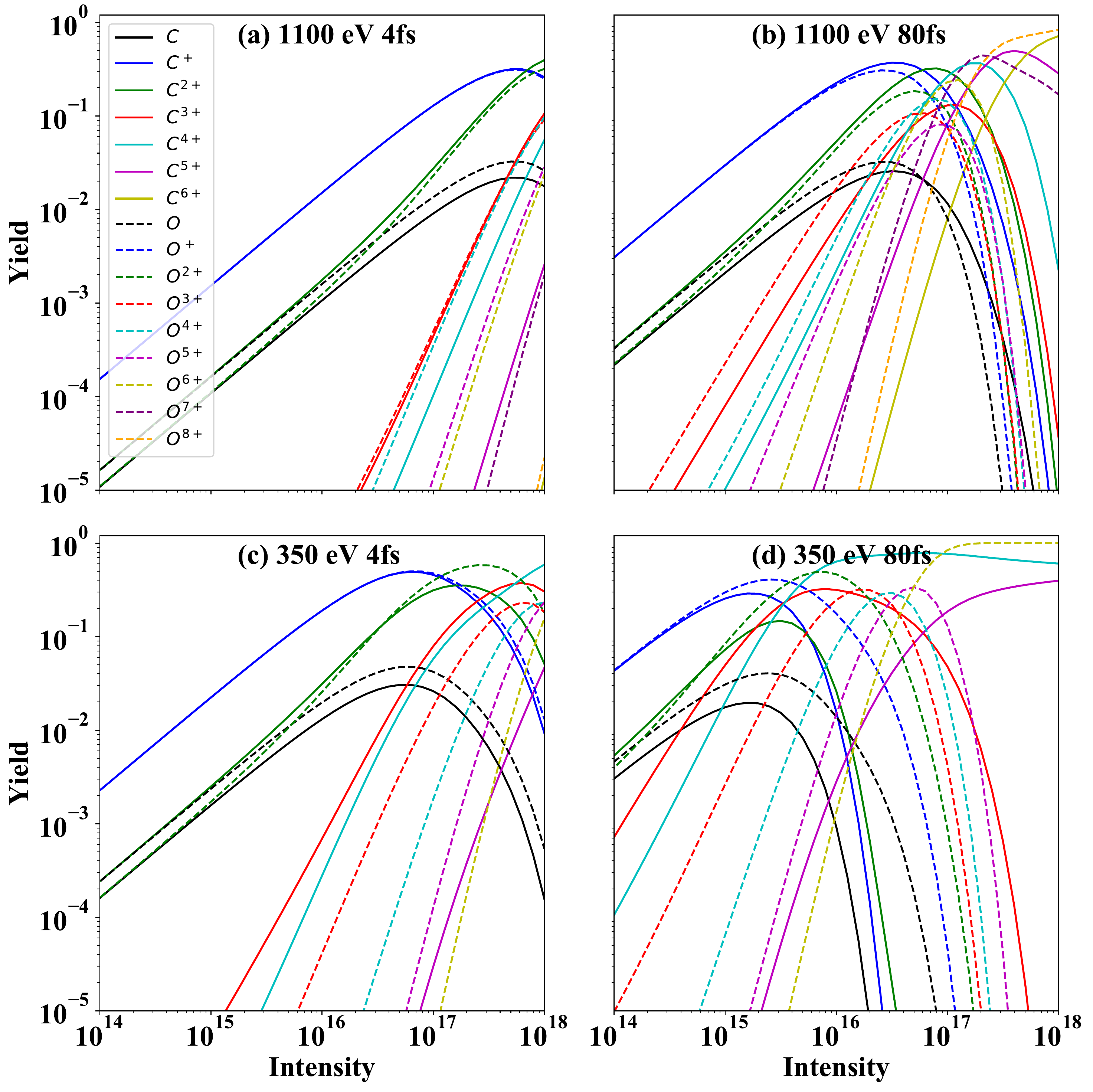}
 }
\caption{Carbon and oxygen ion yields, as a function of intensity, produced by CO interacting with FEL pulses with (a) 1100 eV and 4 fs FWHM (b) 1100 eV and 80 fs FWHM (c) 350 eV and 4 fs FWHM (d) 350 eV and 80 fs FWHM. }
\label{COAtomYields_All}
\end{figure}

In what follows, we identify how the C and O ion yields depend on different FEL pulses interacting with $\rm CO$. In \fig{COAtomYields_All}, we show the atomic ion yields of C and O produced by FEL pulses interacting with $\rm CO$. These ion yields are obtained as a function of intensity for FEL pulses of photon energy 1100 eV and pulse duration 4 fs \fig{COAtomYields_All}(a) and 80 fs \fig{COAtomYields_All}(b) as well as of photon energy 350 eV and pulse duration 4 fs \fig{COAtomYields_All}(c) and 80 fs \fig{COAtomYields_All}(d). The photon energy 1100 eV is sufficient to ionise an electron from the $1\sigma$ orbital, the innermost orbital of $\rm CO$, while 350 eV allows for an electron to be removed from the $2\sigma$ orbital. Therefore, the photon energy of 350 eV does not allow for an electron to be removed from the O site of $\rm CO$. An 80 fs duration FEL pulse allows for more photoionisations transitions to take place compared to a 4 fs duration pulse. 

At a given intensity, for 1100 eV photon energy, comparing \fig{COAtomYields_All}(a) with \fig{COAtomYields_All}(b) and for 350 eV photon energy, comparing \fig{COAtomYields_All}(c) with \fig{COAtomYields_All}(d), we find that as expected the longer 80 fs pulse results in larger yields for the higher-charged states compared to the shorter 4 fs pulse. This is due to a larger number of photons being absorbed during the longer duration pulse. Moreover, we identify the effect of the photon energy, for a given pulse duration, by comparing the ion yields of the 350 eV pulses with the ion yields of the 1100 eV pulses. That is, we compare \fig{COAtomYields_All}(a) with \fig{COAtomYields_All}(c) and \fig{COAtomYields_All}(b) with \fig{COAtomYields_All}(d). We find that, at a given intensity, the ion yields for the higher-charged states are larger at a given intensity for the smaller 350 eV photon energy pulse. This is attributed to two factors. For a given intensity, a lower photon energy corresponds to a higher photon flux. In addition, the photoionisation cross-sections are higher for the lower 350 eV photon energy compared to the higher 1100 eV photon energy, resulting in more photoionisation processes.

\subsubsection{C versus O atomic ion yields}

\begin{figure}
\resizebox{\textwidth}{!}{
 \includegraphics{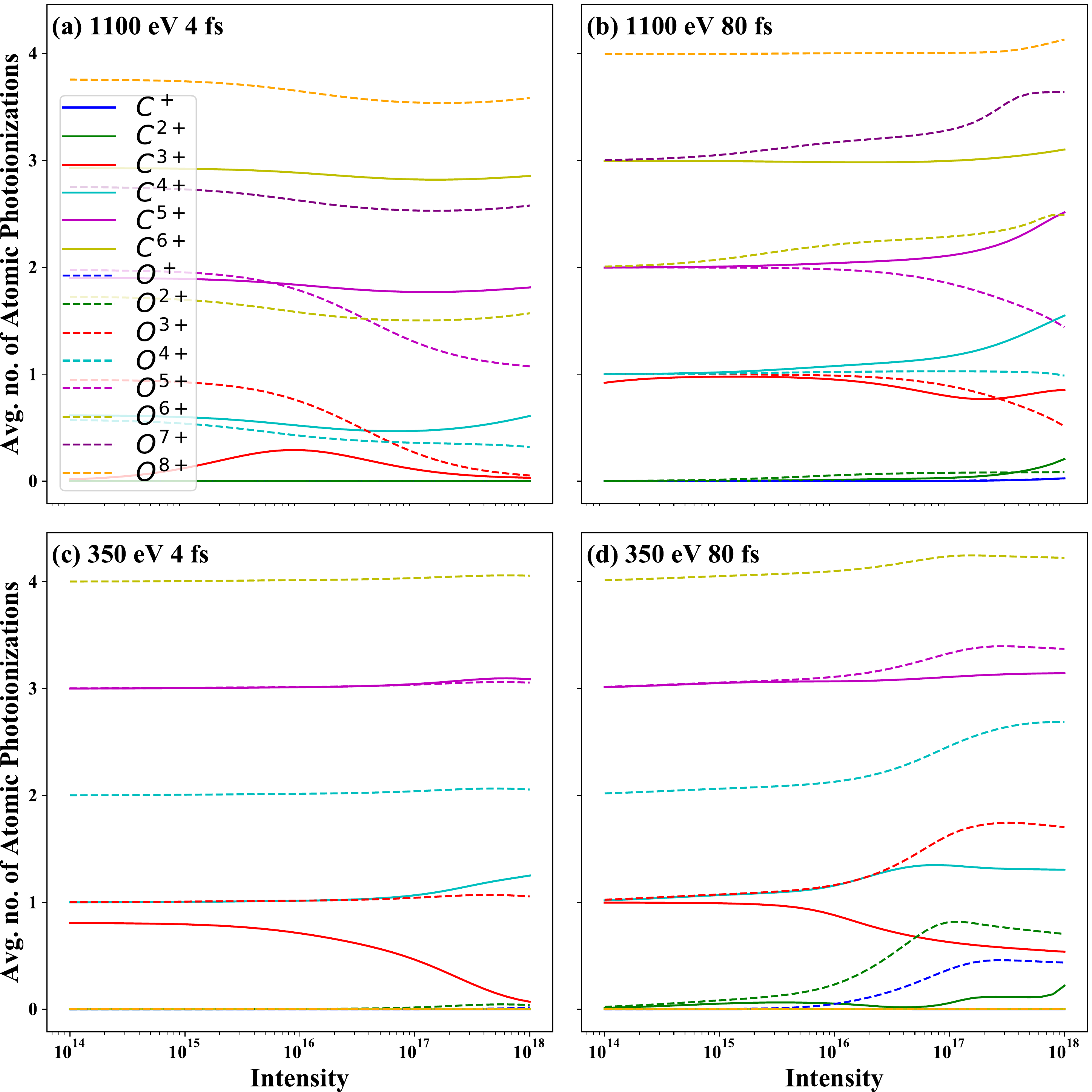}
 }
\caption{Average number of atomic single-photon ionisations, as a function of intensity, for each charged ion produced by CO interacting with FEL pulses with (a) 1100 eV and 4 fs FWHM (b) 1100 eV and 80 fs FWHM (c) 350 eV and 4 fs FWHM (d) 350 eV and 80 fs FWHM.}
\label{COTransYields_All}
\end{figure}

For a given charge state, the C and O atomic ion yields as a function of intensity are generally different, see \fig{COAtomYields_All}. For both 1100 eV and 350 eV photon energies and 4 fs and 80 fs pulse durations, we compare $\rm C$ with $\rm O$, $\rm C^{+}$ with $\rm O^{+}$ and $\rm C^{2+}$ with $\rm O^{2+}$. We find these ion yields to be similar, since they are created mostly by the dissociation of $\rm CO^{2+}$. Specifically, $\rm C^{+}$ and $\rm O^{+}$ are created in equal amounts, see \eq{C+O+Dissociation}. For small intensities, the atomic ion $\rm C^{2+}$ has a slightly higher ion yield than $\rm O^{2+}$, since $\rm C^{2+}$ is produced with a dissociation rate of $0.09\times\Gamma_{a,b}^{\rm CO^{2+}}$ (\eq{C2+Dissociation}), while $\rm O^{2+}$ is produced with a rate of $0.06\times\Gamma_{a,b}^{\rm CO^{2+}}$ (\eq{O2+Dissociation}). For higher intensities, the biggest difference between $\rm C^{2+}$ and $\rm O^{2+}$ ion yields, occurs for pulse parameters 350 eV photon energy and 80 fs pulse duration. The reason is that significantly more atomic photoionisation transitions after dissociation lead to the formation of $\rm O^{2+}$ compared to $\rm C^{2+}$. This is shown in \fig{COTransYields_All}(d). Note, in \fig{COTransYields_All}(a-d), we plot, as a function of intensity, the average number of atomic single-photon ionisation transitions that lead to the formation of each atomic ion state.

We now focus on the higher-charged states $\rm C^{n+}$ and $\rm O^{n+}$, where $n=3,4,5$. At a given intensity, we find that for the 1100 eV photon energy, both for the 4 fs and 80 fs pulses, the yield of the $\rm O^{n+}$ ion is higher than the yield of the $\rm C^{n+}$ ion. The reason is that the single-photon ionisation cross-section to remove an electron from the $1\sigma$ molecular orbital or from the $1s$ orbital in oxygen is higher than the cross-section to remove an electron from the $2\sigma$ molecular orbital or from the $1s$ orbital in carbon. \fig{COTransYields_All}(a) and \fig{COTransYields_All}(b) show that atomic photoionisation transitions play a more important role for the formation of $\rm C^{5+}$ and $\rm O^{5+}$, roughly two transitions, compared to one or less atomic transitions leading to the formation of $\rm C^{n+}$ and $\rm O^{n+}$ with $n=3,4$. At a given intensity, we find that for the 350 eV photon energy, both for the 4 fs and 80 fs pulses, the yield of the $\rm C^{n+}$ ion is higher than the yield of the $\rm O^{n+}$ ion where $n=3$ or $4$. The reason is that a photon energy of 350 eV is insufficient to ionise a core electron corresponding to the oxygen atomic site in CO or to ionise a 1s electron from an oxygen atomic ion. Hence, while $\rm C^{4+}$ is formed by an inner-shell atomic photoionisation from $\rm C^{2+}$ followed by an Auger process, $\rm O^{4+}$ is formed by two valence atomic photoionisation transitions from $\rm O^{2+}$. Indeed, comparing \fig{COTransYields_All}(c-d) with \fig{COTransYields_All}(a-b), we find that the number of atomic transitions to form $\rm C^{4+}$ remains roughly equal to one both for 1100 eV and 350 eV. However, for $\rm O^{4+}$ the number of photoionisations increases to two in the 350 eV case compared to one in the 1100 eV case. Finally, for the 350 eV case, $\rm C^{5+}$ and $\rm O^{5+}$ are both created by three valence atomic photoionisation transitions, resulting in similar ion yields.

\subsubsection{Comparison of C, O and N atomic ion yields}

\begin{figure}
\resizebox{\textwidth}{!}{
 \includegraphics{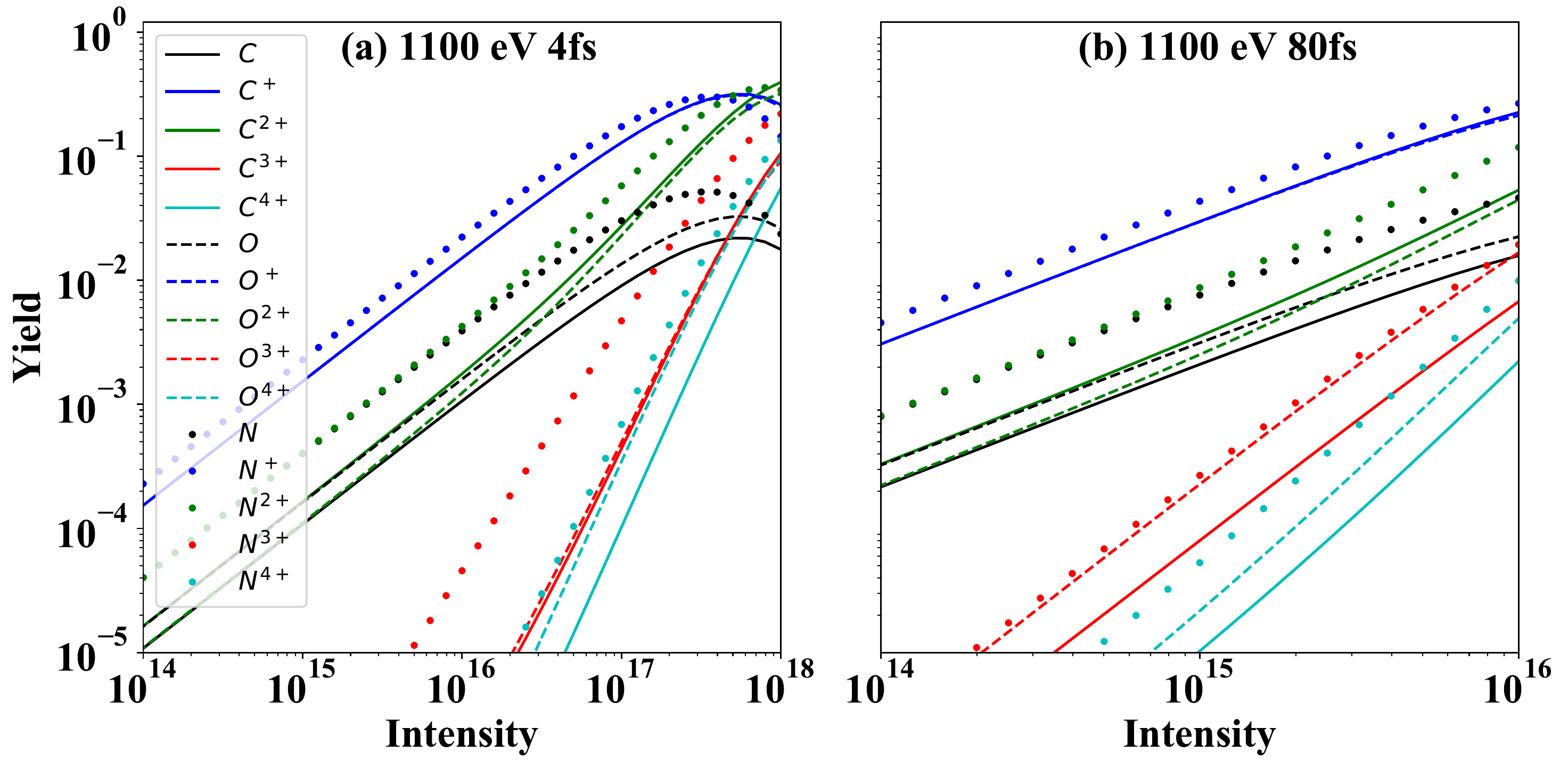}
 }
\caption{Carbon and oxygen ion yields, as a function of intensity, produced by CO interacting with an FEL pulse of 1100 eV photon energy and 4 fs (a) and 80 fs (b) FWHM contrasted with nitrogen ion yields produced by $\rm N_2$ interacting the same pulses.}
\label{CO_N2_AtomYields}
\end{figure}

%{{\bf first you have to describe what people see in the figures and then to interpret what is going on and connect it to the previous figures}}
%To understand the differences between the carbon and oxygen ion yields, we display the average number of atomic transitions that the atomic populations undergo in \fig{COTransYields_All}. Comparing (a) and (b), we can see a large difference in the prevalence of atomic transitions in the 4 fs case and the 80 fs case. In the 80 fs case, it is more likely for dissociation to occur during the pulse, giving time for atomic photoionisations to occur. Looking at the difference between the C and O transitions in (b), we can see why the ion yields diverge more significantly in the 80 fs case of \fig{COAtomYields_All}. We can see that, at the lower photon energy in (c) and (d), there are more atomic transitions at lower intensities for both pulse durations. However, there are no A-O transitions as the oxygen core hole is not accessible at the 350 eV photon energy. For the longer pulse, we see that there are more photoionisations than Auger transitions, whereas for the short pulse this only occurs at high intensities. This is due to the longer pulse allowing for atomic photoionisation transitions before the pulse is over, even when most of the molecular population is dissociating slowly from the $\rm CO^{2+}$ state. For the short pulse, these transitions are only possible when the molecular population is dissociating instantaneously from the $\rm CO^{4+}$ state.

Next, we compare the N atomic ion yields produced by the interaction of $\rm N_2$ with an FEL pulse with the C and O atomic ion yields produced when the same FEL pulse interacts with CO. Specifically, we compare the N, C and O atomic ion yields for an FEL pulse of 1100 eV photon energy and duration 4 fs \fig{CO_N2_AtomYields}(a) and 80 fs \fig{CO_N2_AtomYields}(b). \fig{CO_N2_AtomYields}(a) and  \fig{CO_N2_AtomYields}(b) clearly show that, at a given intensity, each N ion yields is larger than the respective C and O ion yields. This is due to the cross-sections of photoionisation transitions in $\rm N_2$ being higher than the cross-section for photoionisation transitions in CO. Indeed, in $\rm N_2$ the core orbitals $1\sigma_g$ and $1\sigma_u$ have very similar ionisation energies, roughly equal to 680 eV, and large photoionisation cross-sections approximately equal to 0.0023 a.u. for photon energy of 1100 eV. However, in CO, the ionisation energy of the core orbital $1\sigma$, 544 eV, is significantly higher than the ionisation energy of the $2\sigma$, 296 eV. As a result, the photoionisation cross-section from the $1\sigma$ orbital (0.0018 a.u.) in CO is significantly higher than the photoionisation cross-section from the $2\sigma$ orbital (0.00077 a.u.).

\subsection{DCH contributions}

%\iffalse %%%%%%%%%%%MULTI E START

\begin{figure*}
\resizebox{\textwidth}{!}{
 \includegraphics{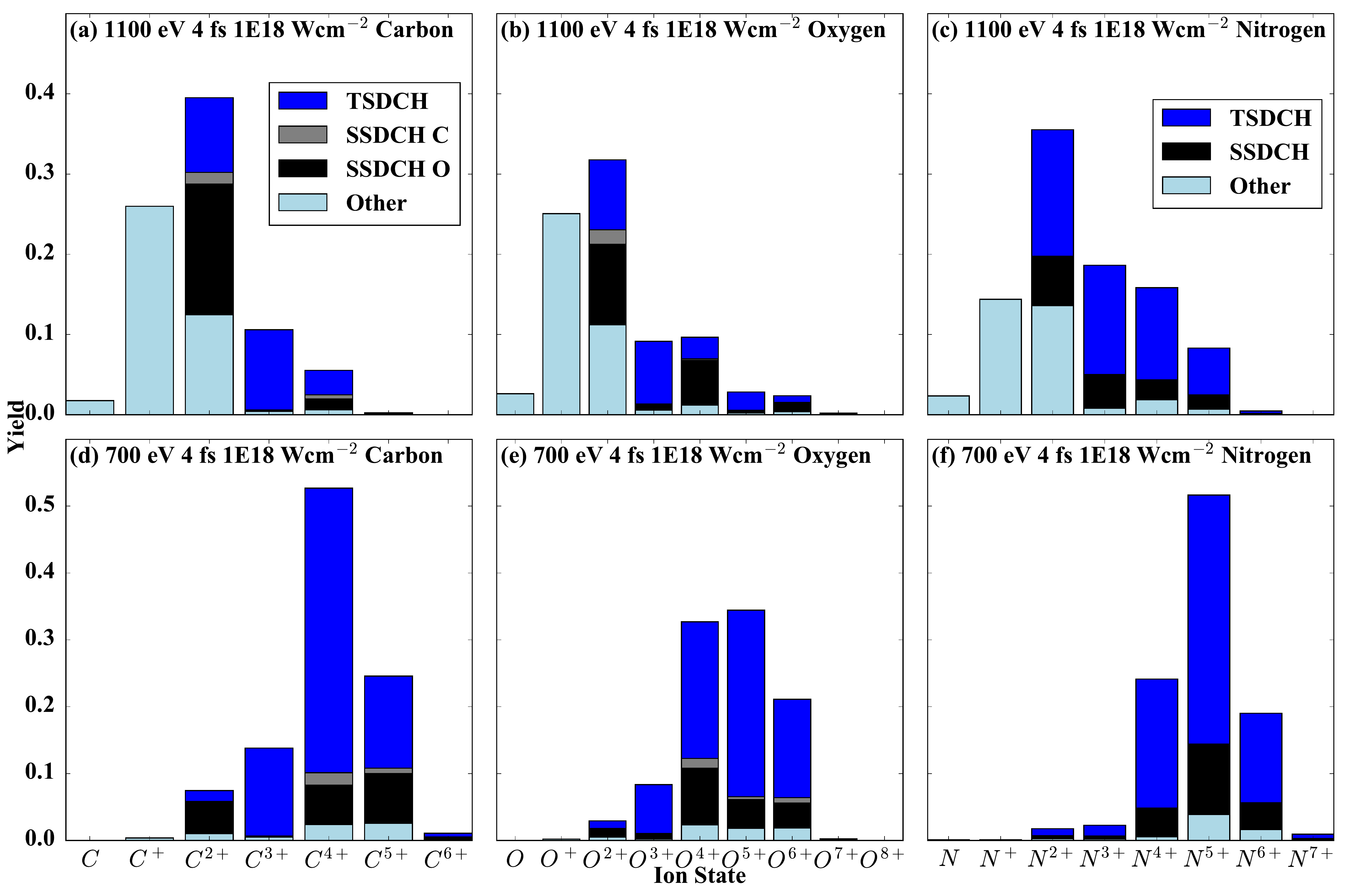}
 }
\caption{Atomic ion yields produced by FEL pulses with 1100 eV or 700 eV photon energy, 4 fs FWHM and $10^{18}$ Wcm$^{-2}$ intensity interacting with CO ((a), (b), (d) and (e))  or $\rm N_2$ ((c) and (f)). For each ion yield, we show the proportion of this ion yield that is reached by accessing a certain type of double-core-hole state.}
\label{IY_CO_N2_DCH}
\end{figure*}

%\fi %%%%%%%%%%%MULTI E END

We are interested in the proportion of each atomic ion yield that is reached by accessing various types of double-core-hole (DCH) states. We calculate these proportions by using an expanded set of rate equations.  That is, for each electronic configuration, we consider four equations, which keep track of the population that reaches this state after accessing a TSDCH state, after accessing a SSDCH state, on the C or on the O site respectively, as well as the population which does not access a DCH state. The proportions of each atomic ion yield that are formed via accessing different DCH states are shown in \fig{IY_CO_N2_DCH} for a 4 fs pulse duration with intensity of $10^{18}\,Wcm^{-2}$. We choose these pulse parameters as they favour the production of DCH states. Indeed, FEL pulses of short duration and high intensity favour more single-photon ionisation transitions occurring in a certain time interval compared to longer and lower intensity FEL pulses. Moreover, in addition to the 1100 eV photon energy used in our calculations in the previous sections, we also consider a photon energy of 700 eV. The reason is that 700 eV is sufficient to photoionise both the $1\sigma$ and $2\sigma$ molecular orbitals in CO, as well as the $1\sigma_g$ and $1\sigma_u$ orbitals in $\rm N_2$ and at the same time these cross-sections are higher than for the 1100 eV case.

%and (a-c) 1100 eV photons (d-f) 700 eV photons. The 700 eV photon energy was chosen to maximise the proportion of TSDCH states produced, as it allows all core photoionisations, but with higher cross-sections due to the lower energy. Similarly, the 4 fs duration and the $10^{18} Wcm^{-2}$ intensity were chosen to promote sequential core photoionisations without intervening Auger or dissociative transitions. 

Examining \fig{IY_CO_N2_DCH}, we see that the proportion of the ion yields which accesses a DCH generally increases at higher-charged ion yields. This is expected as transition pathways which involve a pair of two core photoionisations typically lead to higher-charged ions. For this high-intensity and short-duration FEL pulse the C and O ions with charges 2,3 and 4 are mainly produced by the dissociation of $\rm CO^{4+}$. We find that almost all of the $\rm C^{3+}$ and $\rm O^{3+}$ ion yields are produced via pathways involving TSDCH states. The reason is that after dissociation of $\rm CO^{4+}$ to $\rm C^{2+}$ and $\rm O^{2+}$, each doubly-charged ion is created with a core hole. Each core hole is then filled in by an Auger decay leading to the production of $\rm C^{3+}$ and $\rm O^{3+}$. A large proportion of the $\rm O^{4+}$ yield accesses a SSDCH state with the core holes localised on the oxygen. Regarding the $\rm O^{4+}$ ion, it is formed by dissociation of $\rm CO^{4+}$ to $\rm O^{2+}$ with two $1s$ core holes, which are then filled in by two Auger transitions. Hence, $\rm O^{4+}$ is mostly formed by accessing a SSDCH state of CO. In addition, we find that $\rm C^{2+}$ and $\rm O^{2+}$ are formed after the dissociation of $\rm CO^{4+}$ with no core holes. However, $\rm CO^{4+}$ before dissociation, accesses mostly SSDCH states on the oxygen side. This is due to the much higher photoionisation cross-section to transition from the $1\sigma$ orbital compared to transitioning from the $2\sigma$ orbital. 

Comparing \fig{IY_CO_N2_DCH}(d) with \fig{IY_CO_N2_DCH}(a) and \fig{IY_CO_N2_DCH}(e) with \fig{IY_CO_N2_DCH}(b), we find that, for the 700 eV case, higher-charged ion states are produced with a higher proportion of these ion states accessing a TSDCH state. This is in accord with higher photoionisation cross-sections from both the $1\sigma$ and $2\sigma$ molecular orbitals for 700 eV photon energy, compared to 1100 eV. Finally, comparing the N ion yields in \fig{IY_CO_N2_DCH}(c) with the C and O ion yields in \fig{IY_CO_N2_DCH}(a) and \fig{IY_CO_N2_DCH}(b) for 1100 eV and the N ion yields in \fig{IY_CO_N2_DCH}(f) with the C and O ion yields in \fig{IY_CO_N2_DCH}(d) and \fig{IY_CO_N2_DCH}(e) for 700 eV, we find that $\rm N_2$ dissociates  into higher-charged ion states. Moreover, we find that a higher proportion of each of these N higher-charged ion states accesses a TSDCH state. This is due to the higher photoionisation cross-sections of the $\rm N_2$ core orbitals, particularly the $1\sigma_u$ cross-section, which is much higher than the $2\sigma$ cross-section of $\rm CO$.

\section{Conclusions}

In this work, we have investigated the interaction of FEL pulses with $\rm CO$, a heteronuclear diatomic molecule. In particular, we have calculated the atomic ion yields produced when neutral $\rm CO$ is exposed to a variety of different FEL pulses. We identify higher yields for oxygen ion states for charges 3, 4 and 5, compared to carbon. We also find this to be the case for a photon energy of 1100 eV, which is sufficient to ionise the $1\sigma$ molecular orbital corresponding to the $1s$ core hole on the O site. However, for a photon energy of 350 eV, which does not access the $1\sigma$ molecular orbital, we find that the higher-charged C atomic ions are favoured over the O ions. Finally, we find that high-intensity short-duration laser pulses, with a photon energy sufficient to ionise both core orbitals in CO, favour the production of higher-charged states that are mainly formed by accessing TSDCH states.

{\it Acknowledgments.} A.E. and H. I. B. Banks acknowledge the use of the Legion computational resources at UCL. This work was funded by the Leverhulme Trust Research Project Grant 2017-376.

%%%REFERENCES%%%
%\bibliographystyle{abbrv} %the RSC's .bst file
\bibliography{AugerSub} %You need to replace "rsc" on this line with the name of your .bib file

\end{document}